\documentclass[twocolumn,preprintnumbers,amsmath,amssymb]{revtex4}
\usepackage{graphicx}
\usepackage{xspace}
\usepackage{color}
\newcommand{\ie}{\latin{i.e.}\@\xspace}
\newcommand{\etal}{\latin{et~al}.\@\xspace}
\newcommand{\latin}[1]{\emph{#1}}

\newcommand{\flabel}[1]{\label{fig:#1}}

\newcommand{\Fref}[1]{Fig.~\ref{fig:#1}}

\newcommand{\htilde}{\tilde{h}}
\newcommand{\stilde}{\tilde{s}}

\newcommand{\tautilde}{\tilde{\tau}}

\newcommand{\GC}{\mathcal{G}}

\usepackage{epstopdf}

\newcommand{\beq}{\begin{equation}}
\newcommand{\eeq}{\end{equation}}
\newcommand{\beqn}{\begin{eqnarray}}
\newcommand{\eeqn}{\end{eqnarray}}

\begin{document}

\title{Comment on ``Anomalous Discontinuity at the Percolation Critical
Point of Active Gels''}
\author{Gunnar Pruessner}
\email{g.pruessner@imperial.ac.uk}
\affiliation{Imperial College London,  South Kensington Campus, London SW7 2AZ, United Kingdom}
\author{Chiu Fan Lee}
\email{c.lee@imperial.ac.uk}
\affiliation{Imperial College London, South Kensington Campus, London SW7 2AZ, United Kingdom}
\date{\today}

\maketitle

In their recent work Sheinman \etal introduce a variation of percolation
which they call no-enclaves percolation (NEP). 
The main claims are 1) the salient physics captured in NEP is closer
to what happens experimentally; 2) The Fisher exponent of NEP, is
$\tau=1.82(1)$; 3) Due to the different Fisher exponent, NEP constitutes a
universality class distinct from random percolation (RP).
While we fully agree with 1) and found NEP to be a very interesting variation of random percolation, we disagree with 2) and 3). 
We will
demonstrate that $\tau$ is exactly $2$, directly derivable from 
RP, and thus there is no foundation of a new universality class.

\begin{figure}[bt]
\includegraphics[width=0.8\linewidth]{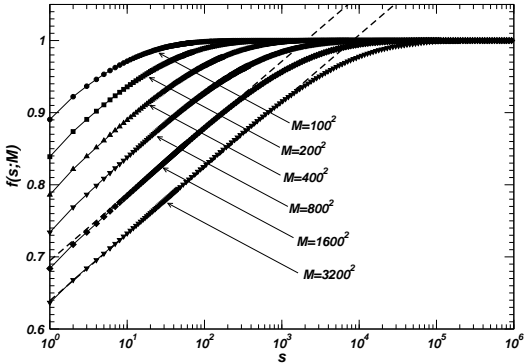}
\caption{\flabel{f_log} Log-linear plot of the fraction $h(s;M)/\htilde(s;M)=f(s;M)$ shown
(binned) for
$M=100^2,\ldots,3200^2$, which is merely a logarithmic
correction; Full lines to guide the eye. 
The dashed lines are linear approximations, 
$1.285+0.04\ln(s/M)$.
}
\end{figure}

In \cite{SheinmanETAL:2015} NEP-clusters are introduced as the union of
all RP clusters with their largest enclosing RP cluster.  It
is argued that their histogram $n_s$, that
is the average number of NEP-clusters of size $s$ (number of sites) 
per realisation of a 
$d$-dimensional system with
$M$ sites, is given by
$n_s \sim M^{\tau-1} s^{-\tau}$
with exponent $\tau<2$. Strictly, 
this cannot
be the scaling form of a finite system, because it lacks a cutoff or
scaling function. Assuming standard finite size scaling
\cite{Barber:1983}, one may
redefine 
$n_s = M^{\tau-1} s^{-\tau} \GC(s/s_c)$,
with
scaling function $\GC(x)$ and
cutoff $s_c$, which requires further
qualification, because it leaves the exponent $\tau$
undefined, unless the choice of $\GC(x)$ is constrained. One
may demand $\lim_{x\to0}\GC(x)=\GC_0$ positive and finite, which renders
$\tau$ the \emph{apparent} exponent \cite{ChristensenETAL:2008}, that is
visible as the slope of the intermediate asymptote in double logarithmic
plots, provided the scaling function is essentially constant. This is
extremely difficult to ascertain and to test
\cite{ClausetShaliziNewman:2009,DelucaCorral:2014}. 

An unambiguous and commonly used definition of
the scaling exponent $\tau$ is
$h(s;M) = s^{-\tau} \GC(s/s_c(M))$
for $s\gg s_0$, the lower cutoff,
demanding that all dependence on the cutoff $s_c(M)$ (which includes finite
size) is contained inside the scaling function. With this definition,
the classical Fisher exponent is recovered in finite size scaling.

We will now use 
that equation
to characterise the scaling of the
\emph{site-normalised} histograms, as normally used in percolation
\cite{StaufferAharony:1994}, of the following models:
$h(s;M)=n_s/M$ (NEP),
$\htilde(\stilde;M)$ (nested NEP or nNEP to be introduced below) and
$h'(s';M)$ (RP),
having exponents $\tau$, $\tautilde$ and $\tau'$ respectively.

The RP site-normalised histogram $h'(s';M)$ follows  $s'^{-\tau'} \GC'(s'/M^{d_f/d})$  where $\tau'=187/91$  is the
Fisher exponent in $d=2$ dimensions and the
clusters'
fractal dimension $d_f$ is  $91/48$ \cite{StaufferAharony:1994}. 

Next, we introduce nNEP, which consists of
clusters from RP but with all of the interior filled up, \ie the
histogram $\htilde(\stilde;M)$ counts cluster sizes $\stilde$, which are
the total areas inside the outermost perimeter of
every RP cluster \cite{CardyZiff:2003}.
 Because nNEP clusters are compact, their size $\stilde$
is related to that of their ``hosting'' RP cluster of size $s'$ by
$\stilde=s'^{d/d_f}$, so that their histogram 
$\htilde(\stilde;M)$ follows $(d_f/d) \stilde^{-2} \GC'( (\stilde/M)^{d_f/d})$,
using
$-\tau' d_f/d - 1 + d_f/d=-2=-\tautilde$.

By construction, nNEP differs
 from NEP only by the latter discounting
all but the largest cluster of a set of nested clusters.  To construct $h(s;M)$ of NEP from $\htilde(s;M)$ of nNEP, we consider the fraction of
those not discounted, $f(s;M)=h(s;M)/\htilde(s;M)$, \ie the 
fraction of those clusters which feature in both histograms. In the
presence of scale invariance $f(s;M)$ is necessarily a function only of
$s/M$, so that $f(s;M)=F(s/M)$, and
can therefore be absorbed into the scaling function,
$h(s;M)=\htilde(s;M)F(s/M)=(d_f/d)s^{-2}
\GC'((s/M)^{d_f/d})F(s/M)$. This \emph{proves} that
$\tau=\tautilde=-2$ exactly. To determine the value of the
\emph{apparent}
exponent, we study $f(s;M)$ directly and find that it is merely a very
small, logarithmic correction, as shown in \Fref{f_log}, which is a log-linear
plot with a remarkably small range at the ordinate. For small arguments,
$f(s;M)$ follows $a+b\ln(s/M)$, before reaching unity and staying there
for $s/M\gtrsim 0.04$; It is clearly not a power law.

In summary, NEP is RP with a different observable, which is not the
order parameter, thus allowing for a ``discontinuity'' in the
approach of the transition. It is an important
model to understand recent experimental work \cite{AlvaradoETAL:2013}, but it does not represent
a novel universality class. The exponent $\tau$ was strongly
underestimated in \cite{SheinmanETAL:2015} due to an unsuitable definition, and if  measured correctly is
exactly
$2$.

\bibliography{articles,books}
\end{document}